\newcommand{\version}{April 3, 2007}

\documentclass[12pt]{article}
\usepackage{bbm}
\usepackage{a4,amsthm,amsfonts,latexsym,amssymb}

\swapnumbers 

\theoremstyle{plain}
\newtheorem{thm}{THEOREM}

\newtheorem{lem}[thm]{LEMMA}

\newtheorem{proposition}[thm]{PROPOSITION}

\newtheorem{fnote}[thm]{FOOTNOTE}
\newtheorem{definition}[thm]{DEFINITION}

\newcommand{\beq}{\begin{equation}}
\newcommand{\eeq}{\end{equation}}
\def\beqa{\begin{eqnarray}}
\def\eeqa{\end{eqnarray}}

\newcommand{\R}{{\mathbb R}}
\newcommand{\Z}{{\mathbb Z}}

\date{\small\version}

\begin{document}
\markboth{\scriptsize{TDS BB  \version}}{\scriptsize{TDS BB
\version}}

\title{
\vspace{-80pt}
\begin{flushright}
{\small UWThPh-2007-09} \vspace{30pt}
\end{flushright}
\bf{Thermodynamic Stability -- A note on a footnote in Ruelle's
book}}
\author{\vspace{8pt} Bernhard Baumgartner$^1$  \\
\vspace{-4pt}\small{Institut f\"ur Theoretische Physik, Universit\"at Wien}\\
\small{Boltzmanngasse 5, A-1090 Vienna, Austria}}

\maketitle

\begin{abstract}
Thermodynamic stable interaction pair potentials which are not of the form
``positive function + real continuous function of positive type''
are presented in dimension one. Construction of such a potential in dimension two
is sketched. These constructions use only elementary calculations.
The mathematical background is discussed separately.
\\[10ex]
\\[3ex]

\end{abstract}

\footnotetext[1]{\texttt{Bernhard.Baumgartner@univie.ac.at}}

\maketitle


\section{Introduction}\label{intro}

In Ruelle's book \cite{R69} on statistical mechanics, in section 3.2
concerning one species of classical particles in $\R^\nu$, you can read:

\begin{proposition}\label{ruelle}
If the pair potential $\Phi$ can be written in the form
\beq\label{sum}
\Phi = \Phi_1 +\Phi_2
\eeq
where $\Phi_1$ is positive, and $\Phi_2$ is a real continuous function
of positive type, then $\Phi$ is stable.
\end{proposition}

``Positive''  is meant here and throughout this paper as nowhere negative,
``stable'' means
\beqa
\exists E_0 \in \R \quad{\rm such}\,{\rm that}\quad \forall N,\,\, \forall \{x_1 ...x_N\}\subset\R^\nu:\quad
U(x_1 \cdots x_N)\,\,\geqq\,\, N\cdot E_0 ,\\
{\rm where}\qquad U(x_1 \cdots x_N)= \sum_{i \neq j} \Phi (x_j-x_i).
\eeqa
This proposition is accompanied by the
\begin{fnote}
It seems to be an open problem
to construct a stable potential which is not of the form (\ref{sum}).
\end{fnote}

We solve this problem in dimension 1, considering particles
either in $\Z$ or in $\R$, giving a detailed proof.
In dimension 2 the problem can also be solved, but we give only a sketch
of the ideas.
\footnote{
Construction in higher dimensions is still an open problem.}

To make it simple, we consider only pair potentials
which are bounded continuous functions and state the stability property as

\begin{definition}\label{stab}
A bounded continuous real valued function $V$ on $\R ^\nu$ is stable, if
\beq\label{stabequ}
E(\rho):=\int\int\rho(x)V(x-y)\rho(y)d^\nu x\,d^\nu y\geq 0
\eeq
for every positive finite measure $\rho(x)d^\nu x$ on $\R ^\nu$.
A bounded real valued function $V$ on $\Z ^\nu$ is stable, if
\beq\label{stabequ2}
E(\rho):=\sum_{\vec{m}}\sum_{\vec{n}}\rho({\vec{m}})V({\vec{m}}-{\vec{n}})\rho({\vec{n}})\geq 0
\eeq
for every positive bounded function $\rho(\vec{m})$ on $\Z ^\nu$.
\end{definition}

The stability property used in Ruelle's Theorem is an immediate
consequence. With $\rho=\sum_{i=1}^N\delta(x_i-x_j)$ put into equation (\ref{stabequ})
one gets
$$U(x_1 \cdots x_N)=E(\rho)-N \cdot V(0)\geq-N \cdot V(0).$$

The main result of our considerations is stated as
\begin{thm}\label{dimone}
Each of the following functions is a stable pair potential,
but not a sum of a positive and a real valued positive definite function.
\begin{enumerate}
\item\label{one}
The function $V:$ $\Z \rightarrow \R$, defined as
\beqa\label{vinzet}
V(0)=V(2)=V(-2)=1,\quad V(1)=V(-1)=-1, \\ V(n)=0\quad\forall n \quad{\rm with}\quad |n|\geq 3,\nonumber
\eeqa

\item\label{two}
The function $W:$ $\R \rightarrow \R$, defined as
\beq\label{wdef}
W(x)=\sum_{n\in\Z} \int_{-\infty}^\infty V(n) f(n-x+y)f(y)dy,
\eeq
with $f$ a positive continuous function $(-\frac12 ,\frac12)\rightarrow \R$ and $V$ as defined in (\ref{vinzet}).

\end{enumerate}
\end{thm}

\section{Properties of the interaction potentials}\label{pot}

\begin{proof} {\it Of part  (\ref{one}) of \ref{dimone} Theorem.}

Denote the distribution of particles on the chain by the ``density'' $\rho$,
a function $\Z\rightarrow \Z_+$.
The interaction energy $U$ becomes smaller, when the system is cut into non-interacting pieces:
If $\rho (n) \geq\rho (n+1)$ divide the chain, cutting between $n+1$ and $n+2$.
Moving the pieces apart, one looses the energy
$$ 2[\rho (n)-\rho (n+1)]\rho (n+2)+2\rho(n+1)\rho(n+3)\geq 0.
$$
The symmetric procedure of cutting between $n-2$ and $n-1$
lowers the energy if $\rho (n-1) \leq\rho (n)$.

Now there remains a set of pieces of no more than three lattice points,
with densities like $$0\leq \rho (n-1) \leq\rho (n) \geq\rho (n+1)\geq 0.$$
Including the ``self-energies'' $N\cdot V(0)$
one gets for each piece, centered around $n$,
$$
E=\rho (n-1)^2+\rho (n)^2+\rho (n+1)^2+2[\rho (n-1)\rho (n+1)-\rho (n-1)\rho (n)-\rho (n)\rho (n+1)]$$
$$=[\rho (n-1)-\rho (n)+\rho (n+1)]^2\geq 0.
$$
Proving the stability of $V$.

If $V$ were the sum of a positive and a positive definite function,
it would give
\beq\label{test}\sum_n V(n)\mu (n)\geq 0,\eeq
for each $\mu$ being both positive and positive definite.
Now consider
\beq\label{mu}
\mu (5\nu)=1,\quad \mu (5\nu \pm 1)=\frac {\sqrt{5}-1}2,\quad  \mu (5\nu \pm 2)=0,
\eeq
which is obviously positive.
Positive definiteness is seen by using Bochner's theorem \cite{RN55}
and calculating the Fourier-Transform, with $\alpha \in(-\pi ,+\pi]$:
\beqa\label{fmu}
\hat{\mu}(\alpha)&=&\sum_n \mu (n) e^{-in\alpha}\nonumber\\
&=&\frac{2\pi}5 \left[ \sqrt{5}\,\delta(\alpha)+\frac{5-\sqrt{5}}2 \left(\delta(\alpha-\frac{2\pi}5),
+\delta(\alpha+\frac{2\pi}5)\right) \right] >0.
\eeqa
But it does not give a positive value in (\ref{test}):
$$
\sum_n V(n)\mu (n)=2-\sqrt{5}\,<\,0.
$$
\end{proof}

The appearance of the numbers $5$ and $\sqrt{5}$
may seem mysterious. Demystifying is the next section, where we present the ``origin''
of these $V$ and $\mu$.

In this section we develop further use of these functions in $\R$ and in $\R^2$.
\begin{proof} {\it Of part (\ref{two}) of \ref{dimone} Theorem.}

For $N$ particles at ${x_1\ldots x_N}$ consider the measure
\beq
\rho (x)= \sum_j \delta (x-x_j).
\eeq
Adding the self-energies $N \cdot W(0)$, we study
\beqa
E&=&\int\int \rho (x)W(x-y)\rho (y)\, dx dy\nonumber\\
&=&\sum_n V(n)\int \rho_f (x+n)\rho_f(x)\,dx,\label {rhonew}
\eeqa
with $\rho_f(x):=\int f(x-y)\rho(y)\,dy$.
Splitting the integral in (\ref{rhonew}) into pieces of intervals with unit length
and defining $\rho_{f,x}(m)=\rho_f(x+m)$ gives
$$
E=\sum_{m\in \Z}\int_0^1dx\sum_n V(n)\rho_f(x+m+n)\rho_f(x+m)$$
$$=\int_0^1dx\sum_{p\in\Z}\sum_{m\in\Z}\rho_{f,x}(p)V(p-m)\rho_{f,x}(m)\geq 0,
$$
by part (\ref{one}) of the theorem.
So the potential $W$ is stable.

Now consider the distribution
\beq\label{mud}
\mu_D(x)=\sum_m\mu (m)\delta(x-m),
\eeq
using the sequence $\mu$ defined in (\ref{mu}).
This distribution is positive and positive definite, as can be seen
at its Fourier transform, which is (up to a factor)
the same as in (\ref{fmu}), now with $\hat{\mu}_D(\alpha +2\pi)=\hat{\mu}_D(\alpha)$
periodically extended to all $\alpha \in \R$.
This $\mu_D$ is used to show that the potential is not a sum of positive and positive definite functions:
\beqa
&\int W(x)\mu_D(x)dx&\nonumber
\\&=&\sum_nV(n)\sum_m\mu (m)\int_{-\frac12}^{+\frac12}dy\int_\R dx\,\delta(x-m)f(n-x+y)f(y)\nonumber
\\
&=&\sum_nV(n)\mu(n)\cdot\int f^2(y)dy\quad <\quad 0.\label{supp}
\eeqa
In the last step the final support of $f$ is essential.
\end{proof}

Construction of a stable pair potential in $\R^2$ being a function of the particle distances only
may be done in the following way:
\begin{itemize}
\item
Use $W(x)$ defined in (\ref{wdef}), now with an $f$ supported on $(-\frac14,\frac14)$,
convolute it twice with the distribution
$$h(x)=\sum_{n\in\Z}e^{-\epsilon |n|}\delta (x-5n):$$
$$W_1(x)=\int\int h(x-y)W(y-z)h(z)dy\,dz.$$
\item
Take the mean value (times $2\pi$) of all rotated versions:
$W_r(\vec{x})=\frac1r W_1(|\vec{x}|).$
\item
Smoothen out $W_r$ with a positive continuous function $g(r)$ with
support on $[0,\frac14)$:
$$W_2(\vec{x})=\int\int g(|\vec{x}-\vec{y}|)\frac{W_1(|\vec{y}-\vec{z}|)}{|\vec{y}-\vec{z}|}
g(|\vec{z}|)d^2y\,d^2z.$$
\end{itemize}

That the stability is not destroyed by the double convolution with $h$
follows from a consideration as it is used in the equation (\ref{rhonew}).
Written in a formal way:
$$\langle\rho|\,W_1\,|\rho\rangle
=\langle\rho|\,h\ast W\ast h_-\,|\rho\rangle=\langle\rho\ast h|\,W\,|\rho\ast h\rangle.$$
Considering only smooth densities $\rho(\vec{x})$ one may take
$W_1(x_1)\delta(x_2)$ as a stable distribution in $\R^2$:
$$\langle\rho|\,W_1\cdot\delta\,|\rho\rangle_{dim=2}
=\int\langle\rho_y|\,W_1\,|\rho_y\rangle_{dim=1}\, d\,y\geq 0.$$
Now rotating the axes and taking the mean value does not destroy the stability.
Once more a double convolution is done, now with $g$ in order to get $W_2$
as a bounded continuous potential acting in $\R^2$.
$$\langle\rho|\,W_2\,|\rho\rangle
=\langle\rho|\,g\ast W\ast g_-\,|\rho\rangle=\langle\rho\ast g|\,W\,|\rho\ast g\rangle\geq 0.$$
Smoothing by convolution with $g$ enables to
consider again sets of particles represented by delta-functions in $\rho$.

To disprove the possibility of splitting $W_2$ into a sum
of a positive and a positive definite function one may use the $\mu_D$ of equ. (\ref{mud})
embedded into $\R^2$,
$$\mu_D(x,y)=\mu_D(x)\delta(y).$$
Due to the smoothing of $W_r$ by $g$ and due to its decrease given by the decrease of $h$,
the integral $\int W_2\mu_D$ is finite:
$$\int W_2\mu_D(x)\,dx=W_2(0)+2W_2(1)\mu(1)
+2\cdot\sum_{\nu=1}^\infty \sum_{n=-2}^{+2}W_2(|5\nu+n|)\mu(n)$$
The bounded support of $f$ and $g$ is needed here as it was in equ. (\ref{supp}).
The exponential decrease implies
$$W_2(|5\nu+n|)=const.\cdot e^{-5\epsilon\,\nu}\frac1{5\nu}V(n)\cdot\left(1+O(\frac1{\nu^2})\right).$$
The ``const.'' factor involves the integrals over $f^2$ and $g^2$,
the error term $O(\frac1{\nu^2})$ gives the difference between $e^{-5\epsilon\,\nu}/5\nu$
and $e^{-\epsilon\,(5\nu+n)}/(5\nu+n)$.
The summations over $\nu$ and $n$ give
$$\approx 2\cdot const.\cdot\sum_{n=-2}^{+2}V(n)\mu(n)\cdot \log(1/\epsilon)+
O\left(\sum_{\nu=1}^\infty e^{-5\epsilon\,\nu}\frac1{\nu^2}\right).$$
The first part is negative and increases without limit when $\epsilon\rightarrow 0$,
while the other term remains finite. So $W_2$ with small $\epsilon$ can {\bf not}
be a sum of positive and positive definite functions.

\section{Mathematical background}\label{math}

Only in applying Proposition \ref{ruelle} in statistical mechanics the
Thermodynamic Limit is considered, not yet in the investigations of ``stability''.
Moreover, in the reformulation in \ref{stab} Definition
there is no mentioning of ``particles''.
What is used of properties of space are:
A distance relation between points and an invariant measure.
This allows for a more general version of the  definition,
concerning functions on groups.
We keep the notation we used above:
$x$ and $y$ are elements of the group, their ``group product'' is $x+y$,
the ``inverse'' of $x$ is $-x$.

\begin{definition}\label{grst}
Consider a bounded continuous real valued function $V$
on a locally compact abelian group $G$ which has the Haar measure $dx$.
$V$ is \emph{stable}, if
\beq
\langle\rho|V|\rho\rangle := \int\int \rho(x)V(x-y)\rho(y)dx\,dy\geq 0
\eeq
for every finite positive Borel measure $\rho(x)dx$.
\end{definition}

Stable functions can be added, multiplied by positive numbers, and limits may be formed.
So they form a closed convex cone, which we call STB.
This cone STB contains POS, the cone of positive functions, also PDF, the cone of
positive definite functions and sums thereof.
\beq\label{vcones}
{\rm STB}\supset {\rm POS}+{\rm PDF}
\eeq
An investigation of the relations between these cones may proceed via investigation of the dual cones
(see \cite{V64, R62, G03}).
The dual cones are subsets of $\mathcal{V}\,'$, the space of finite Borel measures $\mu(x)dx$,
which is the dual space to $\mathcal{V}$, the Banach space of bounded continuous functions.
The dual cone to POS is POS$'$, the set of finite positive Borel measures,
dual to PDF is PDF$'$, the set of finite positive definite Borel measures.
The cone STB$'$ is given as the closure of the cone of convex combinations of
``correlation measures''
\beq
\mu(x)=\int_G \rho(x)\rho(y+x)dy,
\eeq
i.e. convolutions of finite positive Borel measures $\rho(x)dx$
with their reflected version $\rho(-x)dx$.
These correlation measures are both positive and positive definite:
\beq\label{mucones}
{\rm STB'}\subset {\rm POS'}\cap {\rm PDF'}
\eeq

Now the question of equality or inequality in this relation
is related to the central problem which is our concern in this investigation, the
question of equality or inequality in (\ref{vcones}).
If the closed cone ${\rm POS'}\cap {\rm PDF'}$ contains
an element $\mu$ which is not in the closed cone ${\rm STB'}$,
then, by definition of ``dual cone'', there exists an element $V\in {\rm STB}$
such that $\int V\mu <0$, incompatible with a decompostion
$V=f+g,$ $f\in{\rm POS},$ $g\in{\rm PDF}$.

For the groups $\Z_2,\,\Z_3,\,\Z_4$ there is equality in the equations (\ref{vcones}) and (\ref{mucones}),
but not for $\Z_5$.
\begin{proposition}
The intersection of ${\rm POS'}\cap {\rm PDF'}$ with the plane \newline
$\{(\mu(-2)\ldots\mu(2))|\mu(0)=1\}$ is completely characterized by
its extremal points $(0,0,1,0,0)$, $(0,\gamma,1,\gamma,0)$, $(\gamma,0,1,0,\gamma)$,
$(1,1,1,1,1)$, with $\gamma =(\sqrt{5}-1)/2=1/(2|\cos{4\pi/5}|)$.
\end{proposition}
\begin{proof}
By using Bochner's theorem and analyzing the Fourier transform
\beq
\hat{\mu}(k)=\sum_{n=-2}^2\mu(n)e^{-2\pi\, k\,n/5}.
\eeq
\end{proof}
On the other hand there is a bound for STB$'$ which cuts off a
triangular subset of this convex quadrangle:
\begin{lem}
Each element of {\rm STB}$'$ obeys the inequality
\beq
\mu(1)\leq\sum_{n=-2}^2\mu(n)/4.
\eeq
\end{lem}
\begin{proof}
STB$'$ is defined by its extremal rays, formed as correlation measures
of positive densities.
$$\mu\in {\rm STB}',\quad\mu\,\, extremal\quad\Leftrightarrow\quad\exists\rho\geq 0,
\quad\mu(n)=\sum_{n=-2}^2\rho(m)\rho(m+n).$$
Assume, w.l.o.g., that $\rho(-1)\geq\rho(-2)$. then
$$\mu(1)=\left[\rho(-1)+\rho(1)\right]\cdot\left[\rho(-2)+\rho(0)+\rho(2)\right]
-\left[\rho(-1)-\rho(-2)\right]\rho(2)-\rho(-2)\rho(1)$$
$$\leq (\frac s2-x)(\frac s2+x)\leq \frac{s^2}4.$$
Here $s=\sum_m\rho(m)$,\quad $x=\left[\rho(-2)+\rho(0)+\rho(2)-\rho(-1)-\rho(1)\right]/2$.
\newline Observe $\sum_n\mu(n)=s^2$.
\end{proof}
{\bf Remark:}
Also $\mu(2)$ obeys this inequality and $\mu(-1)=\mu(1)$, $\mu(-2)=\mu(2)$.
Closer inspection reveals moreover two rounded edges of STB$'$.

Now the extremal point with $\mu(n)$ as in equation (\ref{mu}) with $\nu=0$
is outside this boundary. And $V(n)$ as in equation (\ref{vinzet})
is an element of STB, but \emph{outside} of POS$+$PDF.

\section{Conclusion}
For pair potentials which are bounded continuous functions the property of being ``stable''
can be reformulated without mention of particles.
In this way it can be studied for abstract abelian groups.
At the heart of the present investigation is the observation of a function $V$ in $\Z_5$
which is stable, but indecomposable into a sum of positive and positive definite functions.
This function $V$ can also be used on $\Z$. With some smoothing
it can be used on $\R$, and in damped periodically extended, rotationally symmetrized
and again smoothed form on $\R^2$.
Of course it is possible find sets of other examples nearby.
So $V(-1)=V(1)$ in Theorem \ref{dimone}
could be a little bit higher than $-1$. Only at $-(\sqrt{5}+1)/4\approx -0.8$
does it become decomposable.

The construction of a rotationally invariant example for dimension two
is not so simple. A nicer one, or one for higher dimension, is not yet known.

\end{document}